\documentclass[pdflatex,sn-mathphys-num]{sn-jnl}

\usepackage{graphicx}%
\usepackage{multirow}%
\usepackage{amsmath,amssymb,amsfonts}%
\usepackage{amsthm}%
\usepackage{mathrsfs}%
\usepackage[title]{appendix}%
\usepackage{xcolor}%
\usepackage{textcomp}%
\usepackage{manyfoot}%
\usepackage{booktabs}%
\usepackage{algorithm}%
\usepackage{algorithmicx}%
\usepackage{algpseudocode}%
\usepackage{listings}%

\theoremstyle{thmstyleone}%
\theoremstyle{thmstyletwo}%

\theoremstyle{thmstylethree}%

\begin{document}

\title[Article Title]{Independent fact-checking organizations exhibit a departure from political neutrality}


\author*[1]{\fnm{Sahajpreet} \sur{Singh}}\email{sahaj.phy@gmail.com}
\equalcont{These authors contributed equally to this work.}

\author*[2]{\fnm{Sarah} \sur{Masud}}\email{sarahm@iiitd.ac.in}
\equalcont{These authors contributed equally to this work.}

\author*[1,3]{\fnm{Tanmoy} \sur{Chakraborty}}\email{tanchak@iitd.ac.in}

\affil[1]{\orgdiv{Department of Electrical Engineering}, \orgname{Indian Institute of Technology Delhi}, \orgaddress{\postcode{110016}, \state{Delhi}, \country{India}}}

\affil[2]{\orgdiv{Department of Computer Science \& Engineering}, \orgname{Indraprastha Institute of Information Technology Delhi}, \orgaddress{\postcode{110020}, \state{Delhi}, \country{India}}}

\affil[3]{\orgdiv{Yardi School of Artificial Intelligence}, \orgname{Indian Institute of Technology Delhi}, \orgaddress{\postcode{110016}, \state{Delhi}, \country{India}}}


\abstract{Independent fact-checking organizations have emerged as the crusaders to debunk fake news. However, they may not always remain neutral, as they can be selective in the false news they choose to expose and in how they present the information. They can deviate from neutrality by being selective in what false news they debunk and how the information is presented. 
Prompting the now popular large language model, GPT-3.5, with journalistic frameworks, we establish a longitudinal measure (2018-2023) for political neutrality that looks beyond the left-right spectrum. Specified on a range of $-1$ to $1$ (with zero being absolute neutrality), we establish the extent of negative portrayal of political entities that makes a difference in the readers' perception in the USA and India. Here, we observe an average score of $-0.17$ and $-0.24$ in the USA and India, respectively. The findings indicate how seemingly objective fact-checking can still carry distorted political views, indirectly and subtly impacting the perception of consumers of the news.}




\maketitle
\section{Introduction}
Independent fact-checking organizations have become paramount in combating the fake-news infodemic \cite{IJoC19728}. Despite their crucial role, the prevalence of media bias in these organizations is yet to be scrutinized. The limitations of left-right binarization of traditional media platforms \cite{doi:10.1073/pnas.2202197119, doi:10.1073/pnas.2013464118, doi:10.1177/0093650215614364} call for a broader perspective on examining the political leaning \cite{Flamino2023}. For example, interview-based studies \cite{Knuutila2022} have established that only 67\% of respondents in the USA and about 55\% of respondents in India are concerned about the risks of fake news. The disparity highlights potential shortcomings in the frameworks for checking fake news \cite{Hoes2024}. Another study \cite{lee2023fact} examined the coherence of fact-checking among PolitiFact and Snopes in the USA and found a 30.4\% disagreement in the final claim. However, this assessment cannot be generalized, given the reliance on an explicit manual labeling system (followed by PolitiFact and Snopes), which is not uniformly available across organizations and geographies. Inspired by these observations and limitations, we attempt to examine how the biases manifest in independent fact-checking organizations. 

Employing journalistic practices of 5W1H (What, Why, Where, When, Who, How) \cite{bleyer1913newspaper} in conjunction with prompting capabilities of large language models (LLMs) \cite{kojima2022large}, we quantify the extent of political neutrality exhibited by these organizations. {A granular continuous scale from -1 to 1 allows us to assess how negatively/positively the entities are projected with the aim of impacting the readers.} The interdisciplinary approach allows a longitudinal investigation (2018-2023) of $\sim24k$ and $\sim10k$ instances from six prominent independent fact-check organizations in the USA and India, respectively. {We hypothesize that the disparity in neutrality among organizations is an amalgamation of topical and narrative accentuations \cite{lee2023fact}.}  

With the inter-organization topical similarity of around $0.8$, our results are analogous to the existing research \cite{lee2023fact}, revealing a selection bias. However, at a microscopic level, the results indicate a higher level of similarity among political entities covered in the debunked news compared to the topic/content of the news. Regardless of geography, all organizations show a deviation from political neutrality. Notwithstanding the clear preference for heads of the states across organizations and geographies, the overall perception meted to entities by an organization remains similar over the five years. All organizations show a deviation from political neutrality -- PolitiFact, Snopes, and Check Your Fact stands at $-0.10$, $-0.28$, and $-0.12$, respectively, while Alt News, Boom, and OpIndia score $-0.28$, $-0.19$, and $-0.25$, respectively.

\section{Materials and Methods}
\textbf{Background.} 
\textcolor{black}{To investigate the extent of selection bias and image portraiture of political entities among the different independent fact-checking websites, we incorporate the content of full-length fact-checking articles published by them. Each article is self-sufficient in terms of the information it conveys about the background of the event, the claims, and the debunked misinformation.}

We enlist prominent independent fact-checking organizations active since $2018$ (Figure \ref{fig:1}(a)) in the USA and India. Details about data curation are provided in Appendix \ref{sec:data}. PolitiFact is exclusively focused on content posted by political entities in the USA, whereas for other fact-checking organizations, no exclusively exists. 
Among the most frequent political entities, we observe Barack Obama and Donald Trump (ex-US presidents), Joe Biden (current US president),  Narendra Modi (current Indian prime minister), Rahul Gandhi (leader of the opposition in India), Arvind Kejriwal (chief minister of the state of Delhi, India), and Yogi Adityanath (chief minister of the state of Uttar Pradesh, India). Here, Modi and Adityanath are associated with the Bhartiya Janta Party (BJP), Gandhi is from the Indian National Congress, and Kejriwal leads the Aam Aadmi Party (AAP). In the USA and India, the president and prime minister are the respective heads of the state.

\textbf{Measuring Topical Overlap.} We prompt GPT-3.5 to answer the ``What" and ``Why" tags of 5W. We further obtain the central claim in the news by employing LLM-based claim normalization \cite{sundriyal2023chaos}. The prompts are enlisted in Appendix \ref{sec:annotation}. We obtain an aggregated embedding per tag per article via the Sentence Transformer \cite{reimers-2019-sentence-bert}. For each tag ($t \in \{What, Why, Claim\}$), we measure the topical similarity ($TS$) between organizations $X$ and $Y$ aggregated ($\Phi[.]$) within a span (``When" tag) of $w=\pm15$ days, with similarity threshold $\tau>0.75$ as summarised in Equation \ref{eq:1}. To reduce the noise in $TS$, $\Phi[.]$ stores only the maximum cosine similarity. For significance testing, we use bootstrap sampling to show a 95\% confidence interval (Appendix \ref{sec:sigtest}).

\begin{equation}
\label{eq:1}
  TS(X^t,Y^t) = \Phi_{i,j}^{|X|\times|Y|}[
Sim(x_i^t,y_j^t)]
\end{equation}

For a given tag $t$, $\forall x_i^t \in X^t = ST(x_i^t)$, where ST is the sentence transformer.

\textbf{Quantifying Neutrality.} We curate a list of the most frequently occurring political entities $E_x$ within an organization (``Who" tag) and access the extent of neutrality ($PS$) meted out by an organization $X$ on entities $e_i \in E_x$ via Equation \ref{eq:2}.

\begin{equation}
\label{eq:2}
    PS(X,e_i) = \frac{|e_{i,j|j=positive}| - |e_{i,j|j=negative}|}{|e_{i,j}|}
\end{equation}

$PS(X,e_i)\approx0$ is deemed neutral, and
$j \in \{positive, negative, neutral\}$ are the labels given to an entity in the article by the organization. We employ the maximum log error to account for the error propagation in $PS$ from LLM. Details about prompting and error estimation of $PS$ are outlined in Appendices \ref{sec:annotation} and \ref{sec:sigtest}, respectively.

\textbf{Data availability.} The dataset and the source code have been made public at \href{https://github.com/sahajps/FC-Bias}{https://github.com/sahajps/FC-Bias}

\section{Results}
\begin{figure*}[t]
\centering
\includegraphics[width=\textwidth]{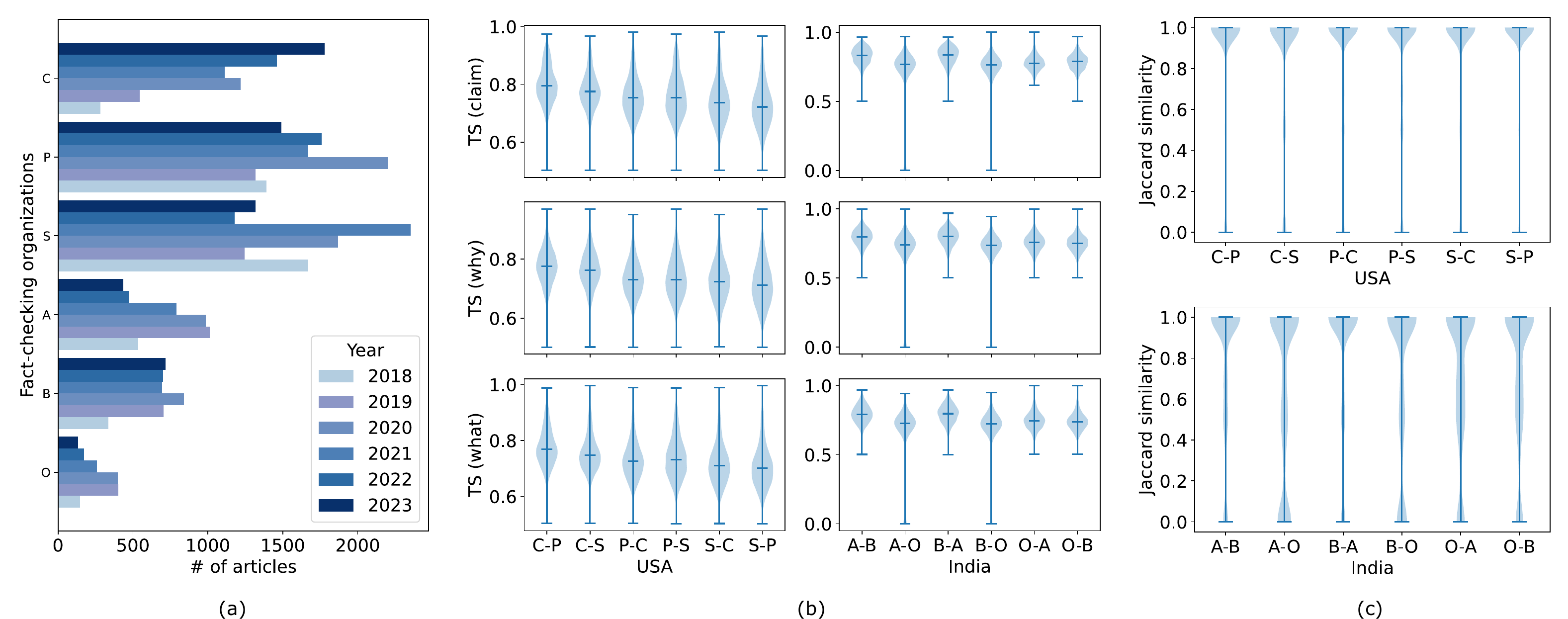}
\caption{The data from six independent fact-checking organizations curated over five years ($2018$-$2023$) -- PolitiFact (P), Snopes (S), and Check Your Fact (C) from the USA, and Alt News  (A), OpIndia (O), and Boom (B) from India. (a) Year-wise breakdown of the number of instances per organization in English. (b) Inter-organization (X-Y) maximum topical similarity (TS) comparing samples of organization X with organization Y within a timeframe of $\pm15$ days and similarity threshold $0.75$. The max TS values are recorded across the ``Claim", 
``What" and ``Why" tags, respectively. (c) Inter-organization (X-Y) Jaccard similarity comparing samples of organization X with organization Y over the $top_k=100$ political entities within a timeframe of $\pm15$ days. 
{Note:} In (b) and (c), we record the similarity only among organizations within a geography.}
\label{fig:1}
\end{figure*}

We note from Figure \ref{fig:1}(b) that the similarity in patterns of topical alignment appears to be consistent over the three tags (``Claim", ``What," and ``Why"), implying that the nucleus of the fake and debunked content is highly coherent within an article. However, within a $15$-day window, the median topical similarity of $<0.8$ indicates that 50\% of articles have poor topical overlap with articles of other organizations. It is corroborated by a narrow width of $<0.01$ obtained when computing the 95\% confidence interval. Interestingly, within the same time window, the overlap among the $100$ most frequently occurring political entities reports a median $\approx 1$\ for the entity-based Jaccard similarity (formula in Appendix \ref{sec:jaccard}). It is further validated by the density distribution in Figure \ref{fig:1}(c). The high entity overlap implies that fact-checking organizations are primarily covering misinformation related to the most talked about political entities, which in turn happen to be heads of state or prominent leaders.

Looking into the frequently occurring entities, we examine the extent of polarity ($PS$) to investigate ``how" the entities are portrayed within an organization. To better illustrate the uncertainty in $PS$, we include maximum log error (formula in Appendix \ref{sec:sigtest}) in Figure \ref{fig:2}. The narrow range of the error bars indicates that even when accounting for maximum deviation, the perception propagated by an organization towards the entities largely remains unaltered. Consequently, it is evident that for both the heads of state, $PS$ is mostly $\geq0$, which may not always align with the portal of their parent political party, even with the same fact-checking organization. 

In terms of leaders of the opposition, all fact-checking organizations in the USA are highly critical of Trump, with an overall $PS$ of $-0.48$ for Check Your Fact, $-0.61$ for Snopes, and $-0.48$ for Politifact against him. Consequently, given the higher negative perceptiveness of the Republican (R) party compared to the Democratic (D) party, we observe an approximate increase in negativity by $6\times$ in Check Your Fact, $7\times$ in Snopes, and $4\times$ in PolitiFact for R/D. Meanwhile, in India, we observe a higher criticality for Rahul Gandhi from OpIndia than by Alt News or Boom. While the neutrality scores against Gandhi are $-0.28$ and $-0.07$ from Alt News and Boom, they reach up to $-0.66$ in OpIndia. OpIndia's preference for the current ruling party becomes more apparent when we see $\approx34\times$ increase in negative sentiment portrayed for Congress vs. BJP by OpIndia. This ratio stands at $0.8\times$ in Alt News and $0.4\times$ in Boom.
 
Since we employ an LLM under a zero-shot setup to provide summarised viewpoints of 5W,  it is imperative to perform a manual quality check. Two expert annotators independently review the samples and provide their assessment on a 0-2 Likert scale. In terms of the relevancy of retrieved tags in TS, we observe an overall average score of What (1.71), Why (1.5), and Claim (1.74). In terms of accessing the recall for entity extraction and the precision of image portrayal, we record an overall average recall of 90.4\% and a precision of 90.2\%. The human evaluation process and the detailed scores are available in Appendix \ref{sec:human_eval}.

\begin{figure*}
\centering
\includegraphics[width=\textwidth]{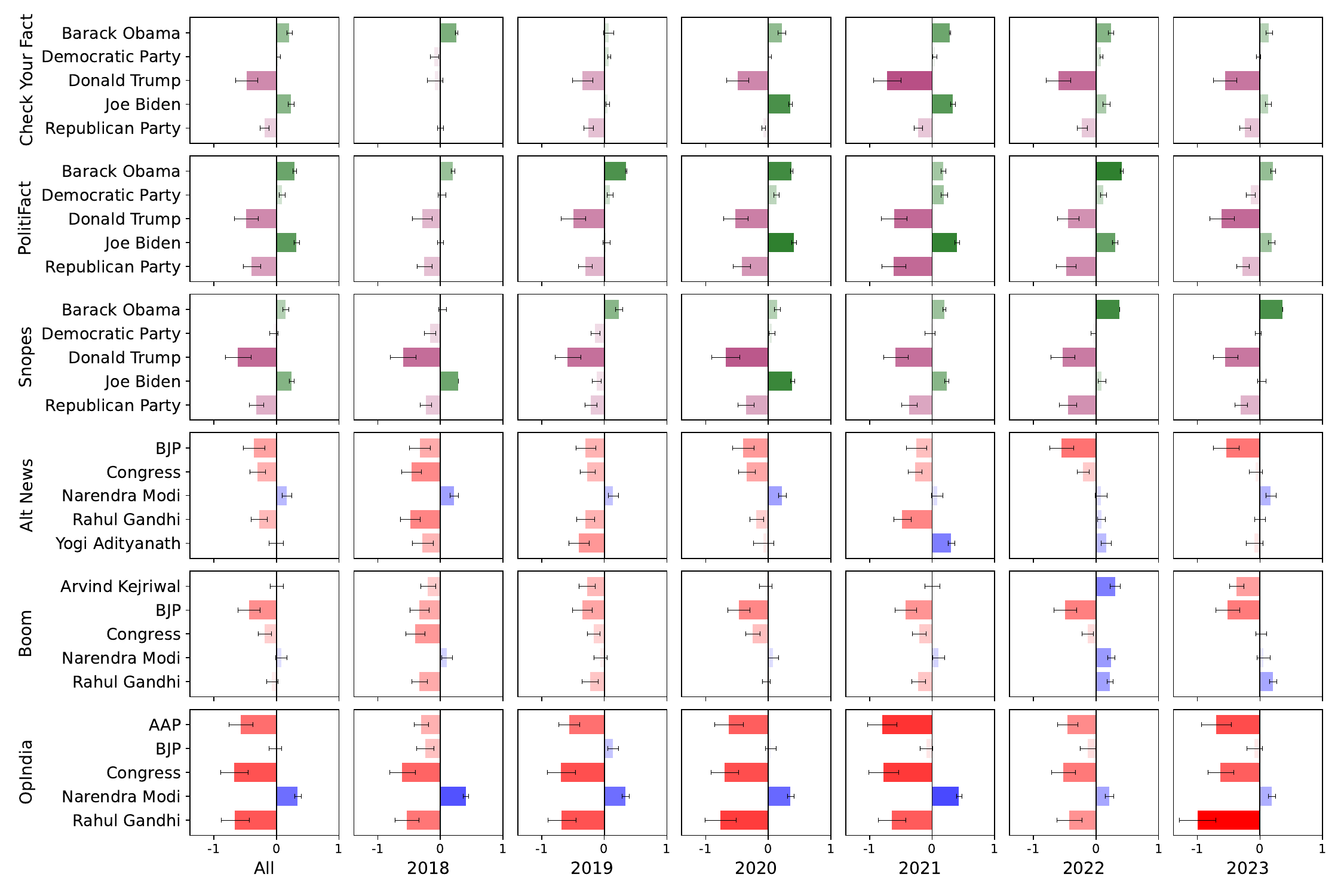}
\caption{The extent of neutrality ($-1\leq PS \leq 1$) for the $top_k=5$ entities per fact-checking organization. PolitiFact, Snopes, and Check Your Fact are in the USA. Alt News, OpIndia, and Boom are based in India. $PS$ establishes ``how" \textit{the coverage of the political entities in the fake news leads to positive, negative, or neutral image for the entity, impacting the reader's perception.} A higher neutrality is observed if $PS\approx0$. Meanwhile, a score closer to -1 (1) highlights a more pessimistic/critical (positive/promoting) tone in terms of portraying the entities. {Note}: The log error bars account for the uncertainty in the prediction of PS.}
\label{fig:2}
\end{figure*}

\section{Discussion}
While the intensity of $PS$ varies from $2018$ to $2023$, the overall sentiment exhibited towards an entity by an organization is more or less fixed, even when accounting for the log error bars. We speculate this arises from both the organization's implicit biases towards these entities as well as the rigidness in entities' ideologies. A more coherent left-right political ideology of the USA vs. India comes into play via compelling dynamics; the person-party alignment in the overall image associated with a leader and their party is better aligned in the USA than in India. Our analysis validates the use of neutrality-based image portrayal, looking beyond the left-right political spectrum \cite{puthillam_kapoor_karandikar_2021}.

It also calls into question the need to quantify political leaning so that fact-checking organizations can be evaluated globally. Having established the robustness of topical similarity (via confidence interval on multiple small subsets), in the future, we can apply the assessments to study the dynamics of specific events around local elections or protests. Notably, the skewness in topical versus entity similarity shows that while the same entities are discussed, ``what" about these entities covered and ``how" they are represented varies. Albeit weak, the skewness and subsequent representation are indicators of selective media bias. By not eulogizing and criticizing political entities equally, fact-checking organizations risk the loss of neutral behavior. 
They exhibit a form of media bias stemming from inequity of coverage rather than a departure from truth \cite{Hamborg2023}. This favoritism portrayal feeds the vicious cycle of jaunced perception among the readers with direct implications on the unassuming users who depend on such organizations for objectivity. We hope our findings empower the readers to become aware of the subtle media biases. Our research prompts future work in understanding consumption patterns for fact-checked news \cite{Kozyreva2024,doi:10.1080/10584609.2019.1668894}.

\backmatter

\bibliography{sn-article}

\begin{appendix}
\section{Data Curation}
\label{sec:data}
While standard media houses also conduct fact-checking, our focus is on organizations dedicated primarily to addressing misinformation and verifying facts, {\em aka} independent fact-checking organizations. From Wikipedia, we shortlist the three most prominent fact-checkers, each from the USA and India -- Check Your Fact\footnote{https://checkyourfact.com}, PolitiFact\footnote{https://www.politifact.com/}, and Snopes\footnote{https://www.snopes.com/} for the USA and Alt News\footnote{https://www.altnews.in/}, Boom\footnote{https://www.boomlive.in/}, and OpIndia\footnote{https://www.opindia.com/} for India due to their higher article counts. In the USA, some fact-checking organizations were notably active before 2018, whereas in India, these websites began gaining prominence around 2018. Therefore, we select the period from 2018 to 2023 for our analysis. 

In order to obtain the fact-checked samples from these organizations, we compose a custom scrapper. At the end of the data curation and filtering process, we obtain
$9,829$ instances from PolitiFact (USA), $9,636$ from Snopes (USA), $6,401$ from Check Your Fact (USA), $4,234$ from Alt News (India), $3,993$ from Boom (India), and $1,520$ from OpIndia (India).

\section{Data Annotation}
\label{sec:annotation}
While these organizations do not blatantly spread misleading information against one political entity (leaders, organizations, official members), at the very least, one can expect the organizations to eulogize and criticize political entities equally, leading to net-neutral political leaning. Given that political leaning and polarity are not explicitly labeled by any independent fact-checking organization, we need to perform large-scale annotations to support our study. To support our analysis, we employ a large language model (LLM) to annotate the samples based on the journalist frameworks of 5W1H (What, When, Why, Who, Where, and How). The 5W1H framework allows us to narrow down the main components of the fake news and the debunked news and nudge the LLM to focus on summarising the central ideas. To further filter out the noise, we leverage the claim normalization \cite{sundriyal2023chaos} capabilities of the LLM. Apart from the 5W1H, we leverage the natural language processing (NLP) techniques of aspect-based sentiment analysis and extrapolate ``positive'', ``negative'' or ``neutral'' sentiment concepts that can be employed to determine the overall image projection/portrayal of the entities mentioned in the fake and debunked news. Throughout the analysis, the LLM employed is GPT-3.5, which we prompted in a zero-shot manner via APIs.

\subsection{Annotations for Topical Similarity}
For each article, we prompt GPT-3.5 with the article's text and direct the system to retrieve the most relevant sentences (``verbatim") from the post in the form of:
\begin{itemize}
    \item \textbf{Prompt I:} \textit{For the given news item, strictly return a list with the most important sentences highlighting the motivation behind spreading the fake news covered in the article. The selected sentences should be the ones that capture the central claim in the fake news. Quote verbatim: do not add any new sentences on your own. Strictly return a JSON decodable Python list format. NEWS ARTICLE: $<$POST$>$.}
    \item \textbf{Prompt II:} \textit{From the following fact-checking news article, quote full and complete sentences from within the post answering the what and why of the 5W based only on the misinformed part and related to the fake news. It is possible that each W can have multiple sentences. Return the result in the format {"what":[], "why":[]}. Quote verbatim: do not add any new sentences on your own; do not paraphrase. Do not focus on how the fact-checking is done. Strictly return a JSON decodable Python list format. NEWS ARTICLE: $<$POST$>$.}
\end{itemize}

\subsection{Annotations for Political Entities and Polarity}
For each article, we prompt GPT-3.5 to retrieve the most relevant political entities (person, organization, or party) along with the associated overall sentiment. For example, if the sentiment is positive, the article helps improve the image of the entity. We employ the following prompt format:
\begin{itemize}
    \item \textbf{Prompt III:} \textit{List (in python dictionary type eg. {a: tag\_a, b: tag\_b}) the names of political figures and parties, categorizing each entity as positive if it helps improve their image, negative if it has the opposite effect, and neutral if it presents balanced views: $<$POST$>$.}
\end{itemize}

\section{Jaccard Similarity for Entity Overlap} 
\label{sec:jaccard}
We take the top 100 entities for each organization based on their frequency of occurrence in articles over the years. These entities are manually annotated to determine whether they are politically relevant. Additionally, we manually map variations of the same entity, such as `President Biden' and `Biden' to `Joe Biden.' Based on the 100 most frequently occurring entities ($E_x$) per organization $X$, we then compute the entity overlap between two organizations $X$ and $Y$ in terms of the Jaccard similarity ($JS$) as outlined in Equation \ref{SIeq:1}. 
\begin{equation}
\label{SIeq:1}
  JS(X,Y) = \frac{|E_x \cap E_Y|}{|E_x \cup E_Y|}
\end{equation}

\section{Significance and Uncertainty Measurement}
\label{sec:sigtest}
This section outlines the process of performing significance testing to corroborate the scores. 

\subsection{Confidence Interval for Topical Similarity}
To assess the significance of the topical similarity between two organizations, $X$ and $Y$, we employ the bootstrap method to calculate the 95\% confidence interval for the median of the topical similarity ($TS$). We generate 10,000 bootstrap samples, with each sample consisting of 20\% of the total number of articles from organization $X$.

\subsection{Maximum Log Error in Polarity Scoring}
The polarity score ($PS$) for various organizations and entities over the years is susceptible to errors in perception-based sentiment label tagging by the LLM. To quantify the maximum propagated error resulting from these prediction inaccuracies, we show the maximum log error as an error bar. Here, Equation \ref{eq:2} can be written in a simple manner (Equation \ref{simple_PS}) for a single entity, where $N_p$ and $N_n$ denote the total number of positive and negative tags, respectively, out of the total occurrence of the entity $N_t$.
\begin{equation}
    PS = \frac{N_p - N_n}{N_t}
    \label{simple_PS}
\end{equation}
Taking natural log and $\Delta$ operator on both sides, we obtain Equation \ref{eq:delta}.

\begin{equation}
\label{eq:delta}
    \frac{\Delta PS}{PS} = \frac{\Delta (N_p - N_n)}{N_p - N_n} - \frac{\Delta N_t}{N_t}
\end{equation}

Note the fact that $N_t$, the total number of occurrences of an entity, is constant when we specifically look at the entity tagging; then the maximum error in $N_t$ is $\Delta N_t=0$. And since we want to calculate the maximum log error, $\Delta (N_p - N_n) = \Delta N_p + \Delta N_n$, the above equation can be reformulated as Equation \ref{eq:delta_nt}.
\begin{equation}
\label{eq:delta_nt}
    \frac{\Delta PS}{PS} = \frac{\Delta N_p + \Delta N_n}{N_p - N_n}
\end{equation}
Here, $\Delta N_{p/n}=N_{p/n} * (1-Precision_{p/n})$, and this can be understood with the intuition that if we take $1-Precision$ as an error in a single instance,  the total error will be multiple of $N_{p/n}$ (Equation \ref{eq:precision1}).
\begin{equation} 
\label{eq:precision1}
    \frac{\Delta PS}{PS} = \frac{N_p*(1-Precision_p) + N_n*(1-Precision_n)}{N_p - N_n}
\end{equation}

Using Equation \ref{simple_PS}, the error difference can be written as Equation \ref{eq:delta_ps} and simplified as Equation \ref{maxlogerr_PS}.
\begin{equation}
\label{eq:delta_ps}
    \Delta PS = \frac{N_p*(1-Precision_p) + N_n*(1-Precision_n)}{N_p - N_n} * \frac{N_p - N_n}{N_t}
\end{equation}

\begin{equation}
    \label{maxlogerr_PS}
    \Delta PS = \frac{N_p*(1-Precision_p) + N_n*(1-Precision_n)}{N_t}
\end{equation}

\section{Human Evaluation of LLM Generated Outputs}
\label{sec:human_eval}
Since we employ an LLM under a zero-shot setup to provide summarised viewpoints via Prompts I, II, and III, it is imperative to perform a thorough quality check of the results before we can generalize them at scale. Here, we employ the help of two expert annotators who are familiar with both American and Indian politics and understand how fact-checking works. The annotators are one male and one female, aged 20-30, who are also familiar with computational social science and prompt engineering in LLMs. The annotators are randomly provided fact-checked news items from the six organizations (with references to the organization anonymized to reduce bias) along with the predictions obtained from prompts I, II, and III.  The annotators independently perform an assessment on $20$ samples for the following:
\begin{itemize}
    \item \textbf{Relevancy score for retrieved What, Why, and Claim:} We ask annotators to score the retrieved sentences for relevance on a Likert scale between $0$ to $2$ with $0$ being irrelevant, $1$ being somewhat relevant, and $2$ being highly relevant. For evaluator 1, the average relevancy scores were 1.75 for What, 1.67 for Why, and 1.87 for Claim, while for evaluator 2, the scores were 1.67 for What, 1.33 for Why, and 1.61 for Claim. 

    \item \textbf{Recall of extracted entities and precision of generated perception tags:} We ask human annotators to manually identify the major political entities in the articles and assign each one a perception tag of positive, negative, or neutral. Based on this, we find the average recall of relevant political entities to be 93.3\% for one evaluator and 87.5\% for another. The average precision (same for both evaluators) of the annotated tags is 90.2\%. Specifically, the precision for positive and neutral classes is perfect, while the precision for the negative class is 70.6\%. We further use this precision to find the uncertainty using maximum log error for the polarity score (refer to Equation \ref{maxlogerr_PS}) due to errors in the LLM's output.
\end{itemize}

\end{appendix}

\end{document}